\documentclass{PoS}

\def\beq{\begin{equation}}
\def\eeq{\end{equation}}
\def\beq{\begin{equation}}
\def\eeq{\end{equation}}
\def\bea{\begin{eqnarray}}
\def\eea{\end{eqnarray}}
\def\nn{\nonumber}
\def\Eq#1{Eq.~(\ref{#1})}
\def\ii{\imath 0}
\def\td#1{\tilde{\delta}\left(#1\right)}

\def\M#1{{\cal M}^{(#1)}} 
\def\MD#1{{\cal M}^{(#1) \, \dagger}} 
\def\nlo{{\rm NLO}}
\def\v{{\rm V}}
\def\r{{\rm R}}
\def\la{\langle}
\def\ra{\rangle}

\def\nn{\nonumber}
\def\Eq#1{Eq.~(\ref{#1})}

\def\td#1{\tilde{\delta}\left(#1\right)}

\title{
\vspace*{-2cm}
\begin{minipage}{\textwidth}
{\normalfont\small IFIC/18-01
\hspace{\fill} January 2018
}\\
\end{minipage}\\[60pt] 
Recent developments from the loop-tree duality}

\ShortTitle{Recent developments from the loop-tree duality}

\author{\speaker{Germ\'an Rodrigo}, F\'elix Driencourt-Mangin \\
        Instituto de F\'{\i}sica Corpuscular, Universitat de Val\`{e}ncia -- 
        Consejo Superior de Investigaciones Cient\'{\i}ficas, Parc Cient\'{\i}fic, E-46980 Paterna, Valencia, Spain.\\
        E-mail: \email{german.rodrigo@csic.es,felix.dm@ific.uv.es}}

\author{Germ\'an F. R. Sborlini\\
        Dipartimento di Fisica, Universit\`a di Milano and INFN Sezione di Milano, I-20133 Milan, Italy. \\
        E-mail: \email{german.sborlini@unimi.it}}

\author{Roger J. Hern\'andez-Pinto\\
        Facultad de Ciencias F\'isico-Matem\'aticas, Universidad Aut\'onoma de Sinaloa, Ciudad Universitaria, 
        CP 80000, Culiac\'an, Sinaloa, M\'exico. \\
        E-mail: \email{roger@uas.edu.mx}}

\abstract{In this talk, we review the most recent developments of the four-dimen\-sional unsubstraction (FDU) and loop-tree duality (LTD) methods. In particular, we make emphasis on the advantages of the LTD formalism regarding asymptotic expansions of loop integrands.}

\FullConference{13th International Symposium on Radiative Corrections\\
		24-29 September, 2017\\
		St. Gilgen, Austria}

\begin{document}

\section{Introduction}
Our present understanding of the fundamental interactions among elementary particles is described through the Quantum Field Theory (QFT) called the Standard Model (SM), and potential extensions. Besides the mathematical and physical beauty of this theory, it also exhibits some unphysical weaknesses that lead to cumbersome technical difficulties. Some of these problems arise because quantum corrections are described by loop diagrams in which the validity of the theory is extrapolated to arbitrary large energy scales, much above the Planck scale. This leads to the appearance of ultraviolet (UV) divergences, which can be solved by absorbing the singular behaviour inside the redefinition of the parameters available in the theory.

On the other hand, there are other singularities whose origin is located in the low-energy regime or associated with the presence of multiple degenerated configurations. In fact, the existence of massless particles that can be emitted with zero energy leads to quantum states which are treated differently from those describing the absence of real radiation. Also, it is possible to emit particles in parallel directions, in such a way that they can not be distinguished from the emission of a single particle. These kinematical configurations translate into the presence of infrared (IR) singularities, which need to be properly handled in order to achieve physically meaningful results.  

In the context of perturbative computations within QFT, it is mandatory to introduce a regularisation framework to explicitly identify all the singularities mentioned above; otherwise it would not be possible to extract any physical prediction from the mathematical formulation. In Dimensional Regularisation (DREG)~\cite{Bollini:1972ui,tHooft:1972tcz,Cicuta:1972jf,Ashmore:1972uj,Wilson:1972cf}, that constitutes one of the customary methods, the number of space-time dimensions is modified from $d=4$ to $d=4-2\varepsilon$, thus rendering the intermediate steps mathematically well-defined. Explicitly, the application of this dimensional shift allows to exhibit all the singularities as poles in $\varepsilon$, once the integration in the loop momenta and/or the phase-space of real radiation has been performed. 

After the removal of the UV singularities through renormalisation, the virtual and the real quantum corrections contribute with poles in $\varepsilon$ whose coefficients have opposite 
signs, thus leading to a finite result. Although this procedure efficiently transforms the theory into a calculable and well-defined mathematical framework, the evaluation of loop integrals and the adequate subtraction of  the singularities of the real radiation in higher space-time dimensions
require a big computational effort, particularly at higher perturbative orders. A great simplification of perturbative calculations would be possible if we could apply a regularisation method directly in four space-time dimensions while preserving most of the symmetries of the original theory, as DREG does. To this extent, and besides different variants of DREG, several groups have defined new regularisation schemes that do not alter the dimensions of the space-time, or change it to an integer number, such as the four-dimensional formulation (FDF)~\cite{Fazio:2014xea} of the four-dimensional helicity scheme, the six-dimensional formalism (SDF)~\cite{Bern:2002zk}, implicit regularisation (IREG)~\cite{Battistel:1998sz}, four-dimensional regularisation / renormalisation (FDR)~\cite{Donati:2013voa}, and the four-dimen\-sional unsubtraction (FDU)~\cite{Sborlini:2016gbr}. A review and comparison of these methods was recently published in Ref. \cite{Gnendiger:2017pys}.

The purpose of this talk is to introduce the FDU framework~\cite{Sborlini:2016gbr,Hernandez-Pinto:2015ysa,Sborlini:2016hat,Driencourt-Mangin:2017gop,Ramirez-Uribe:2017gbf} and to briefly review some of the recent developments. We start by emphasizing the main features of the loop-tree duality theorem (LTD)~\cite{Catani:2008xa,Bierenbaum:2010cy,Bierenbaum:2012th,Buchta:2014dfa,Buchta:2015wna}, which constitutes one of the cornerstones of the whole formalism. Another key ingredient of the formalism is the real-virtual momentum mapping, that allows to perform the summation over the degenerate soft and collinear configurations achieving a fully local cancellation of IR singularities between real and virtual contributions. On the other hand, it is also possible to rewrite the customary UV subtraction counter-terms to get a local cancellation of singularities without altering the number of space-time dimensions. In this way, the FDU method should improve the efficiency of Monte Carlo event generators because it is meant for integrating simultaneously real and virtual contributions. Also, we motivate the advantages of the LTD for carrying out asymptotic expansions of loop integrands; since it reduces the loop integration domain to the Euclidean space of the loop three-momentum, expanding the integrands and integrating leads to the right result straightforward. 

The LTD formalism or a similar framework has also been used to derive causality and unitarity constraints~\cite{Tomboulis:2017rvd}, or to integrate numerically subtraction 
terms~\cite{Seth:2016hmv}. Besides this, it can be related to the forward limit of scattering amplitudes~\cite{Catani:2008xa,CaronHuot:2010zt}. Moreover, the LTD formalism has been applied very recently in the framework of the color-kinematics duality~\cite{Jurado:2017xut}.

\section{Introduction to the loop-tree duality}
One of the main bottlenecks in perturbative computations is related with the calculation of loop integrals. 
In the loop-tree duality \cite{Catani:2008xa,Bierenbaum:2010cy,Bierenbaum:2012th}, any loop integral or loop scattering amplitude is transformed into a sum of tree-level like objects. These objects are built by setting on-shell a number of internal loop propagators equal to the number of loops, thus opening any multi-loop diagram into a tree topology.

The explicit realisation of the LTD is achieved by modifying the $\ii$ prescription of the Feynman propagators that remain off-shell, transforming them into \emph{dual} propagators. For instance, when the internal momentum $q_i$ is set on-shell, we have
\beq
\td{q_i} \, G_D(q_i;q_j) = \td{q_i} \, \frac{1}{q_j^2-m_j^2-\ii \, \eta \cdot k_{ji}}~,
\eeq
with $k_{ji}^{\mu} = q_j^{\mu}-q_i^{\mu}$ and $\eta^\mu$ an arbitrary future-like vector. For the sake of simplicity, we can set $\eta^\mu=(1,\bf{0})$, which is equivalent to integrate out the loop energy component through the Cauchy residue theorem. Thus, the left-over integration is then restricted to the loop three-momentum. The on-shell condition is given by $\td{q_i} = \imath \, 2 \pi \, \theta(q_{i,0}) \, \delta(q_i^2-m_i^2)$, and determines that the loop integration is restricted to the positive energy modes of the on-shell hyperboloids (light-cones for massless particles).

The dual prescription can hence be $+\ii$ for some of the dual propagators, and $-\ii$ for others; this encodes in a compact and elegant way the contribution of the multiple cuts that are 
introduced by the Feynman tree theorem~\cite{Feynman:1963ax}. Moreover, this change of sign in the prescription is responsible for the consistency of the partial cancelation of singularities among different dual contributions. A careful analysis of the singular behaviour of the loop integrand shows that all the physical threshold and IR singularities are restricted to a compact region of the loop three-momentum~\cite{Buchta:2014dfa}, even if the on-shell loop three-momenta are unrestricted. 

\section{Four-dimensional unsubtraction at NLO and beyond}
\begin{figure}[t]
\begin{center}
\includegraphics[width=12.5cm]{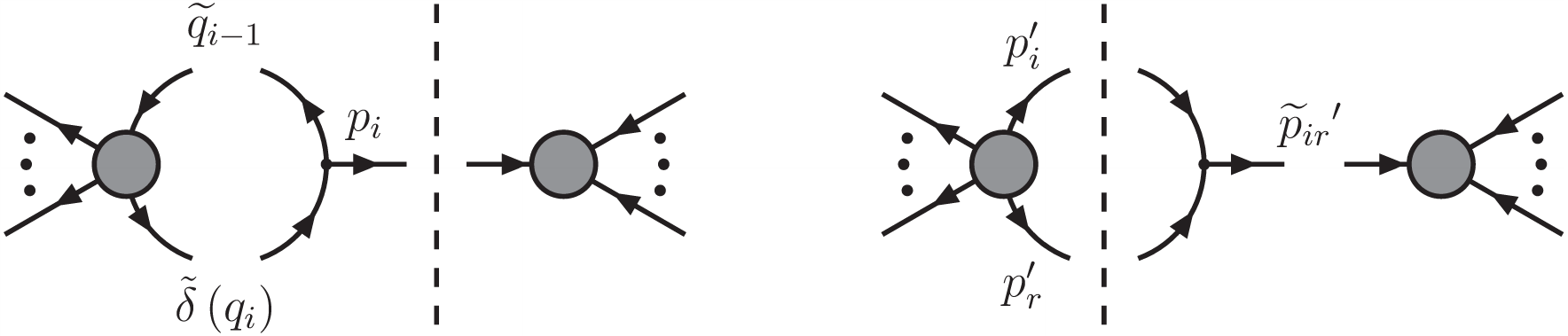}
\caption{\label{fig:collinear}
Interference of the Born process with the one-loop scattering amplitude with 
internal momentum $q_i$ on-shell, $\M{1}_{N}(\td{q_i})\otimes \MD{0}_{N}$ (left), 
and interference of real processes with parton splitting $p_{ir}'\to p_i'+p_r'$:
$\M{0}_{N+1}\otimes \MD{0}_{N+1}(p_{ir}')$ (right). The dashed line represents momentum conservation.
In the soft/collinear limits the momenta $q_{i-1}=q_i-p_i$ and $p_{ir}'$ become on-shell
and the scattering amplitudes factorise.}
\end{center}
\end{figure}

The compactness of the region containing the IR singularities in the loop three-momentum integration domain is \emph{the} key point that allows to build the momentum mappings between the real and the virtual kinematics. As shown in Fig.~\ref{fig:collinear}, it is possible to exploit the factorisation properties of gauge theories (for instance, QCD) to motivate the explicit form of the mapping and perform the summation over degenerate soft and collinear states.

As an starting example, let's explain how the framework applies to the usual NLO cross-section computations. It is a well known fact that, due to the Kinoshita-Lee-Nauenberg theorem, the NLO cross-section is built from the renormalised one-loop virtual correction with $m$ partons in the final 
state and the exclusive real cross-section with $m+1$ partons in the final state, namely
\beq
\sigma^\nlo = \int_{m} d\sigma_{\v}^{(1,\r)}+ \int_{m+1} d\sigma_{\r}^{(1)}~.
\eeq
Eventually, we would need to introduce initial-state IR counter-terms, which would account for the collinear singularities which are absorbed into the parton distribution functions. For the sake of simplicity, we will skip this point in the following discussion. It will be explicitly treated in a 
future publication. 

In the context of FDU, the virtual contribution is obtained from its LTD representation,  
\beq
d\sigma_{\v}^{(1,\r)} = \sum_i \int_\ell \,  2 \, {\rm Re} \, \la \M{0}_N|\M{1,\r}_N(\td{q_i}) \ra \, 
{\cal O}_N (\{p_j\}) ~,
\label{eq:nlov}
\eeq
where $\M{0}_N$ is the $N$-leg scattering amplitude at LO and $\M{1,\r}_N$ is the renormalised 
one-loop scattering amplitude, that also contains the self-energy corrections of the external legs. Notice that self-energies alter the local behaviour of the integrand-level expressions, even if in the massless case they are usually ignored because they lead to vanishing contributions after integration within DREG. In particular, these contributions are responsible for both IR and UV singularities, that might be properly cancelled in order to get integrable functions. Also, appropriate counter-terms that locally subtract the UV singularities must be included in the definition of the renormalised amplitude \cite{Sborlini:2016gbr,Ramirez-Uribe:2017gbf}. Besides the structure of the amplitudes, the delta function inside $\M{1,\r}_N$ symbolises the dual contribution with the internal momentum $q_i$ set on-shell; a sum over all possible single-cuts is required according to the LTD theorem. Finally, notice that the integral is weighted with the function ${\cal O}_N$ that defines a given observable, i.e. it imposes cuts in the phase-space measure and implements the experimental reconstruction algorithms (such as jet or isolation algorithms).

On the other hand, the real contributions have to be rewritten using the same integration variables that appear in the dual terms, i.e. those used to describe the Born phase-space, and the loop three-momentum. Of course, this connection is possible through the definition of a suitable momentum mapping, which is simplified if we isolate all the possible IR singular regions present in the real emission phase-space. Thus, the very first step consists of introducing a complete partition given by
\beq
\sum_i {\cal R}_i (\{p_j'\}) = \sum_i \prod_{jk\ne ir} \theta(y_{jk}' -y_{ir}')   = 1~,
\eeq 
which is equivalent to split the real phase-space as a function of the minimal dimensionless two-body invariants $y_{ir}'=s_{ir}'/s$. So, in general, we have 
\beq
\int_{m+1}  =  \int_{m} \, \int_\ell \, \sum_i  {\cal J}(q_i,\{p_j\})  \, {\cal R}_i(q_i,\{p_j\})~,
\eeq
where ${\cal J}(q_i,\{p_i\})$ is the Jacobian of the transformation $\{p_j'\}={\cal T}_i((q_i,\{p_i\}))$ and ${\cal R}_i(q_i,\{p_j\})= {\cal R}_i(\{p_j'\})$ are the characteristic functions that select each partition.

Combining the dual and the mapped real contributions, the NLO cross-section can be cast in the form
\bea
\sigma^\nlo &=& \int_{m} \int_\ell \,  \sum_i  \bigg[ 2 \, {\rm Re} \, \la \M{0}_N|\M{1,\r}_N(\td{q_i}) \ra \, {\cal O}_N (\{p_j\}) \nn \\
&+& {\cal J}(q_i,\{p_j\}) \, {\cal R}_i(\{p_j'\}) \, |\M{0}_{N+1}(\{p_j'\})|^2 \, {\cal O}_{N+1}(\{p_j'\}) \bigg]~, 
\label{eq:nlovr}
\eea 
where the external momenta $p_j'$, the real phase-space and the tree-level scattering amplitude $\M{0}_{N+1}$ are rewritten in terms of the loop three-momentum (equivalently, the internal on-shell loop momenta) and the external momenta $p_i$ of the Born process through the transformation ${\cal T}_i$ in each partition.

The cross-section defined in \Eq{eq:nlovr} has a smooth four-dimensional limit and can be evaluated directly in four space-time dimensions. This is possible because there is a completely local cancellation of singularities, that renders the real-virtual combination into an integrable function. In this way, there is no need to introduce (or to keep) any additional regularisation framework to carry out the computation. DREG is only necessary to fix the UV renormalisation 
counter-terms in order to define the cross-section within a conventional scheme (for instance, the $\rm{\overline{MS}}$ scheme): the whole computation could directly be implemented numerically in four space-time dimensions even without fixing the scheme, thus completely skipping DREG. 

The FDU at NLO has been successfully applied to compute decay rates in scalar toy models~\cite{Hernandez-Pinto:2015ysa}, and $A^* \to q \bar q(+g)$ with $A=\{H,\phi,Z,W\}$ (for both massless and massive quarks) \cite{Sborlini:2016gbr,Sborlini:2016hat} and to deal with the calculation of one-loop amplitudes involving Higgs production and decay \cite{Driencourt-Mangin:2017gop}. 

The core ideas behind FDU at one-loop can be extended to deal with NNLO computations. Within the usual framework, the total NNLO cross-section consists of three contributions,
\beq
\sigma^{\rm NNLO} = \int_{m} d\sigma_{\v\v}^{(2,\r)} + \int_{m+1} d\sigma_{\v\r}^{(2,\r)} + \int_{m+2} d\sigma_{\r\r}^{(2)}~,
\eeq
where the double virtual cross-section $d\sigma_{\v\v}^{(2,\r)}$ contains the interference of the two-loop with the Born scattering amplitudes and the square of the one-loop scattering amplitude with $m$ final-state particles; the virtual-real cross-section $d\sigma_{\v\r}^{(2,\r)}$ includes the contributions from the interference of one-loop and tree-level scattering amplitudes with one extra external particle; and the double real cross-section $d\sigma_{\r\r}^{(2)}$ are tree-level contributions with emission of two extra particles. 

The LTD representation of the two-loop scattering amplitude is obtained by setting two internal propagators on-shell~\cite{Bierenbaum:2010cy}, and it leads to the two-loop dual components
$\la \M{0}_N|\M{2,\r}_N(\td{q_i,q_j})\ra$. The two-loop momenta of the squared one-loop 
amplitude are independent and generate the dual contributions $\la \M{1,\r}_N(\td{q_i})|\M{1,\r}_N(\td{q_j})\ra$. It is important to notice that, in both cases, there are two independent loop three-momenta and $m$ final-state momenta. On the other hand, the dual representation of the real-virtual contribution  $d\sigma_{\v\r}^{(2,\r)}$ is obtained by setting only one internal propagator on-shell. Thus, they include an additional external momenta and a free loop three-momentum, i.e. two independent integration momenta. Analogously, the double-real terms are associated with the presence of two additional real-particles in the final state. So, in the end, through the definition of a proper momentum mapping, the three contributions can be merged at integrand level, and define an integrable function in four space-time dimensions. More details about the explicit implementation of a two-loop computation inside this framework will be provided in a forthcoming publication~\cite{INPREP2}.

\section{Application to asymptotic expansions}
Computing Feynman integrals for multi-leg processes is in general a complicated task. From the mathematical point of view, one of the reasons for the high complexity of this computations is related with the presence of many different scales, that leads to several possibilities for the functional dependence of the results. To simplify this problem, we can focus in certain kinematical configurations and properly expand the result. In this direction, the asymptotic expansions are useful when we are interested in kinematical configurations exhibiting a hierarchy of physical scales, masses and external momenta, in such a way that the final answer can be approximated by series depending on the ratios of different scales. 

The series expansion of the integrated results is well defined, but the main bottleneck is related with the integration itself. So, it is much more useful to find a suitable expansion at integrand-level, in such a way that:
\begin{itemize}
\item the integration of each individual coefficient is easier to perform, compared with the original full integral;
\item and the sum of the integrated coefficients of the expansion agrees with the expansion of the full integral. This can be summarised by the concept of commutation between integration and series expansion.
\end{itemize}
However, in general, the naive expansion of loop integrands does not lead to the right result. Several complementary expansions in different kinematical regions need to be considered, which constitutes the method of expansion by regions~\cite{Beneke:1997zp,Smirnov:2002pj}. The reason for that is simply due to the fact that in a Minkowski space, such as the loop momentum, the metric does not induce a positive-definite scalar product, i.e. there are non-trivial null vectors satisfying $p \cdot p =0$. Of course, this is not true in the Euclidean space, since the induced scalar product fulfils that if $p \cdot p =0$ then $p^\mu=0$. So, this explains why the LTD formalism could provide an advantage for introducing asymptotic expansions: after integrating out the energy component of the loop momentum, the left over integration lies in the Euclidean space of the loop three-momentum. 

In order to provide a practical example, let's consider the naive integrand-level expansion of the following one-loop integral:
\bea
&& \int_\ell \frac{1}{(\ell^2-M^2+\ii)[(\ell+p)^2-M^2+\imath 0]} \nn \\ && = 
\int _\ell \frac{1}{(\ell^2-M^2+\ii)^2} \left( 1 + \frac{2\ell\cdot p + p^2}{\ell^2-M^2+\ii} + \cdots \right)~,
\label{minko}
\eea
in the case $0< p^2 \ll M^2$ with $p_0>0$. However, this expansion is not valid in the region where $\ell^2 \simeq M^2$ because in that case $p^2$ is not the smallest quantity in the denominator of the second Feynman propagator. Therefore, \Eq{minko} needs to be balanced with the expansion in a complementary kinematical region to obtain the full result. On the other hand, using LTD and analysing the first of the two cuts, we find
\beq
-\int_\ell \frac{\td{\ell}}{2\ell\cdot p + p^2-\ii} = -\int_\ell \frac{\td{\ell}}{2\ell\cdot p}  \sum_{n=0}^\infty \left( \frac{-p^2}{2\ell\cdot p}\right)^n~,
\label{dualexpansion}
\eeq
with $\td{\ell} = \imath \, 2\pi \, \theta(\ell_0) \, \delta(\ell^2-M^2)$. But since $\ell$ is a positive energy vector with mass $M$ (i.e. it fullfils the condition imposed by the on-shell delta), the scalar product $2\ell\cdot p$ is always positive and of $2\ell\cdot p \approx {\cal O}(M)$. Thus, the Taylor expansion in \Eq{dualexpansion} is well defined for any value of the loop momentum. 

In Ref.~\cite{Driencourt-Mangin:2017gop} this procedure has been used, for the first time, to obtain the asymptotic expressions for the $H\to \gamma\gamma$ one-loop amplitude. To test the efficiency of the LTD approach, we obtained the corresponding LTD representation both in the large and the small mass limits of the internal particle running in the loop. As expected, the results satisfied the commutativity between integration and series expansion, and we managed to reproduce the known results. Moreover, the LTD/FDU approach allowed us to identify a universal structure for the $H \to \gamma \gamma$ one-loop amplitudes, i.e. that the functional dependence of the dual representation of the amplitude is independent of the nature of the particles circulating the loop (scalars, fermions or $W$ bosons). We expect that this result holds at the two-loop level~\cite{INPREP2}.

\section{Conclusions}
We have reviewed the FDU/LTD formalism for the calculation of physical cross-sections and differential distributions. Within this approach, all the IR and UV 
singularities are canceled locally, and the virtual and real corrections are evaluated simultaneously. In this sense, we expect an improvement of the numerical efficiency of Monte Carlo implementations, since the integration over the real and the virtual phase-space is performed at once and the integrand exhibits a smooth behaviour (due to the local cancellation of singularities). Besides this, we have shown that LTD is advantageous regarding the direct evaluation of asymptotic expansions of loop integrands because it transforms the integration domain into an Euclidean space.

\section*{Acknowledgements}
This work is supported by the Spanish Government and ERDF funds from European
Commission (Grants No. FPA2014-53631-C2-1-P and SEV-2014- 0398), by Generalitat Valenciana (Grant No. PROMETEO/2017/057),
and by Consejo Superior de Investigaciones Cient\'{\i}ficas (Grant No. PIE-201750E021). FDM acknowledges support from Generalitat Valenciana (GRISOLIA/2015/035) and GFRS from Fondazione Cariplo under the Grant No. 2015-0761. 
The work of RJHP is supported by CONACyT M\'exico.


\end{document}